\pdfoutput =0
\let \origdocument \document 
 
\def \document {\endgroup \endinput }


\title{Optimal Number of Transmit Antennas for Secrecy Enhancement in Massive MIMOME Channels}

\author{
\IEEEauthorblockN{
Saba Asaad\IEEEauthorrefmark{1}\IEEEauthorrefmark{2},
Ali Bereyhi\IEEEauthorrefmark{2},
Ralf R. M\"uller\IEEEauthorrefmark{2},
Rafael F. Schaefer\IEEEauthorrefmark{3},
Amir M. Rabiei\IEEEauthorrefmark{1}
}
\IEEEauthorblockA{
\IEEEauthorrefmark{1}School of Electrical and Computer Engineering, University of Tehran\\
\IEEEauthorrefmark{2}Institute for Digital Communications (IDC), Friedrich-Alexander Universit\"at Erlangen-N\"urnberg (FAU)\\
\IEEEauthorrefmark{3}Information Theory and Applications Chair, Technische Universität Berlin (TUB)\\
saba{\_}asaad@ut.ac.ir, ali.bereyhi@fau.de, ralf.r.mueller@fau.de, rafael.schaefer@tu-berlin.de, rabiei@ut.ac.ir
\thanks{This work was supported by the German Research Foundation, Deutsche Forschungsgemeinschaft (DFG), under Grant No. MU 3735/2-1.}
}
}


\IEEEoverridecommandlockouts

\maketitle

\begin{abstract}
This paper studies the impact of transmit antenna selection on the secrecy performance of massive MIMO wiretap channels. We consider a scenario in which a multi-antenna transmitter selects a subset of transmit antennas with the strongest channel gains. Confidential messages are then transmitted to a multi-antenna legitimate receiver while the channel is being overheard by a multi-antenna eavesdropper. For this setup, we approximate the distribution of the instantaneous secrecy rate in the large-system limit. The approximation enables us to investigate the optimal number of selected antennas which maximizes the asymptotic secrecy throughput of the system. We show that increasing the number of selected antennas enhances the secrecy performance of the system up to some optimal value, and that further growth in the number of selected antennas has a destructive effect. Using the large-system approximation, we obtain the optimal number of selected antennas analytically for various scenarios. Our numerical investigations show an accurate match between simulations and the analytic results even for not so large dimensions.
\end{abstract}

\IEEEpeerreviewmaketitle

\section{Introduction}
\label{sec:intro}
Massive \ac{mimo} systems have been identified as a key technology for the next generation of wireless communication systems \cite{marzetta2010noncooperative}. Consequently, physical layer security of these systems has gained significant attentions in recent years \cite{kapetanovic2015physical}. The main premise in physical layer security is to exploit the inherent characteristics~of wireless channels. The pioneering work of Wyner \cite{wyner1975wire}, considered the wiretap channel as a basic model for secure transmission and demonstrated that secrecy is obtained as long as the legitimate receiver observes a better channel than the eavesdropper. Provisioning secrecy for \ac{mimo}~wiretap channels, also referred to as \ac{mimome} channels, was then studied in the literature widely \cite{khisti2010secure,oggier2011secrecy}. The fact of using multiple antennas in these systems can significantly improve the secrecy performance by means of focusing the transmission beam to the legitimate receiver. This technique in massive \ac{mimo} systems effectively provides secrecy due to the large number of antennas, and makes them robust against passive eavesdropping \cite{kapetanovic2015physical}. 

The performance gains in \ac{mimo} systems are mainly obtained at the expense of elevated complexity and cost. In fact, the growth in the number of antennas, increases both the computational complexity and \ac{rf}-cost significantly. Consequently, addressing solutions to alleviate these issues has become a topic of interest. Antenna selection is introduced as a possible solution which reduces the overall \ac{rf}-cost, as well as the computational complexity, without significant performance loss \cite{molisch2004mimo}. This solution was recently further raised up in the context of massive \ac{mimo} systems, due to the large dimensions of these systems \cite{li2014energy, asaad2017tas}.

Although antenna selection vanishes the robustness of massive \ac{mimo} systems against passive eavesdropping, it still can enhance the secrecy performance compared to conventional \ac{mimo} systems. For single \ac{tas}, the problem of secure transmission was initially studied in \cite{alves2011enhanced}. The authors in \cite{alves2011enhanced} considered a scenario in which a multi-antenna transmitter with a single \ac{rf}-chain intends to send confidential messages to a single-antenna legitimate~receiver while a single-antenna eavesdropper is overhearing the channel. The study was later extended in \cite{alves2012performance} to the case of multi-antenna eavesdropper where the authors showed that, similar to the case with a single-antenna eavesdropper, the secrecy outage probability improves when the number of antennas at the transmitter side increases. \cite{yang2013physical} investigated the \ac{tas} for secure transmission in the general \ac{mimome} wiretap channel assuming that a single \ac{rf}-chain available at the transmitter. The problem was further studied in \cite{wang2014secure} for Nakagami-$m$ fading considering the single \ac{tas}. The impacts of the imperfect channel estimation and antenna correlation were also investigated in \cite{al2017secrecy} and \cite{yang2013physical}. In contrast to the single \ac{tas}, the secrecy performance of massive \ac{mimome} channels under the multiple \ac{tas} has not been yet addressed in the literature. In this case, increasing the number of transmit antennas is beneficial to both the legitimate receiver and eavesdropper simultaneously, and therefore, its effect on the overall secrecy performance is not clear. 
%
%
\subsection*{Contributions}

In this paper, we study the impact of multiple \ac{tas} on~the secrecy performance of massive \ac{mimome} channels. In particular, we consider a general \ac{mimome} setup with a large number of transmit antennas in which a subset of antennas with the strongest channel gains is selected. We show that under the given \ac{tas} protocol, increasing the number of selected antennas up to an optimal value can enhance the secrecy performance of the system. The impact, however, can be destructive if the number of the selected antennas exceeds the optimal value.
\subsection*{Notation}
Throughout the paper the following notations are adopted. Scalars, vectors and matrices are represented with non-bold, bold lower case and bold upper case letters, respectively.~$\mH^{\her}$ indicates the Hermitian of $\mH$, and $\mI_N$ is the $N\times N$ identity matrix. The determinant of $\mH$ and Euclidean norm of~$\bx$~are denoted by $\abs{\mH}$ and $\norm{\bx}$. $\log\left(\cdot\right)$ and $\loge \left(\cdot\right)$ indicate the~binary and natural logarithm respectively, and $\mone_{\set{\cdot}}$ denotes the indicator function.  $\E\set{\cdot}$ is statistical expectation, and
\begin{align}
\rmQ(x)= \int_{x}^{\infty} \phi(x) dx
\end{align}
represents the standard $\rmQ$-function where $\phi(x)$ is the zero-mean and unit-variance Gaussian probability~density~function.

\section{Problem Formulation}
\label{sec:sys}
Consider a Gaussian \ac{mimome} channel in which the transmitter, legitimate receiver and eavesdropper are equipped with  $N_\rmt$, $N_\rr$ and $N_\ee$ antennas, respectively. At each transmit interval, the transmitter encodes its messages into the codeword $\bx_{N_\rmt \times 1}$ and sends it to the legitimate receiver. The received signal is given by the vector $\by_{N_\rr \times 1}$ which reads 
\begin{align}
\by=\sqrt{\rho_\mm} \ \mH_\mm \bx + \bn_\mm. \label{eq:sys-1}
\end{align}
Here, $\rho_\mm$ denotes the average \ac{snr} at each receive antenna, $\bn_{\mm}$ is circularly symmetric complex Gaussian noise with zero-mean and unit-variance, i.e., $\bn_\mm \sim \cnormal{\mI}$ and $\mH_\mm$ is an ${N_\rr \times N_\rmt}$ \ac{iid} unit-variance quasi-static Rayleigh fading channel matrix and is referred to as the main channel. The~ea- vesdropper overhears $\bx$ and receives $\bz_{N_\ee \times 1}$ given by
\begin{align}
\bz=\sqrt{\rho_\ee} \ \mH_\ee \hspace{.2mm} \bx + \bn_\ee, \label{eq:sys-2}
\end{align}
where $\rho_\ee$ is the average \ac{snr} at each of the eavesdropper~antennas, $\bn_\ee \sim \cnormal{\mI}$, and $\mH_\ee$ identifies an $N_\ee \times N_\rmt$~\ac{iid} unit-variance quasi-static Rayleigh fading channel matrix between the transmitter and eavesdropper. We denote it as~the eavesdropper channel. The legitimate receiver and eavesdropper have the~\ac{csi}~of~their channels. At the transmit side, however, the \ac{csi} of the~channels is not necessarily available. Moreover, the main and eave- sdropper channels are supposed to be statistically independent.


\subsection{\ac{tas} Protocol}
\label{sec:tas}
Let the $N_\rr \times 1$ vector $\bfh_{\mm j}$ denote the $j$th column vector of $\mH_\mm$ for $j\in \set{1, \ldots, N_\mathrm{t}}$. Represent the index set~of~order statistics from the arranging of vector norms $\norm{\bfh_{\mm j}}$~in~decreasing order by $\setW\coloneqq\left\lbrace w_1, \ldots, w_{N_\mathrm{t}} \right\rbrace$, i.e., 
\begin{align}
\norm{\bfh_{\mm w_1}} \geq \norm{\bfh_{\mm w_2}} \geq \cdots \geq \norm{\bfh_{\mm w_{N_\mathrm{t}}}}. \label{eq:sys-4}
\end{align}
At each transmission interval, the transmitter selects $L_\rmt$ transmit antennas via the \ac{tas} protocol $\mas$ as follows:
\begin{enumerate}[label=(\alph*)]
\item The ordered index set $\setW$ is determined at the transmitter and legitimate receiver. At the transmit side, the task is done either by supposing the transmitter to estimate the channel gains itself, or assuming $\setW$ to be evaluated by the legitimate receiver and given to the transmitter through a rate-limited return channel.
\item The transmitter then selects the $L_\rmt$ antennas which correspond to the index subset $\setW_\mas\coloneqq \set{w_1, \ldots, w_{L_\rmt}}$ and transmits over them with equal average power.
\end{enumerate}
\begin{remark}
Considering the task of determining $\setW$, the transmitter, even in the absence of a return channel, need not~acquire the complete \ac{csi}. In fact, as $\setW$ is determined via the ordering in \eqref{eq:sys-4}, the transmitter only needs to estimate the channel norms. The task which can be done at the prior uplink stage simply by attaching \ac{rf} power meters at each transmit antenna, and requires a significantly reduced time interval compared to the case of complete \ac{csi} estimation \cite{narasimhamurthy2009antenna}.
\end{remark}

\subsection{Achievable Secrecy Rate}
For the \ac{mimome} channel specified in \eqref{eq:sys-1} and \eqref{eq:sys-2},~the~instantaneous secrecy rate is expressed as \cite{khisti2010secure,oggier2011secrecy}
\begin{align}
\mar_\rms =[\mar_\mm-\mar_\ee]^+\label{eq:sys-R_s}
\end{align}
where $[x]^+=\max\set{0,x}$. Here, $\mar_\mm$ denotes the achievable rate over the main channel which is determined as
\begin{align}
\mar_\mm=\log\abs{\mI+\rho_\mm\mH_\mm \mQ \mH_\mm^\her} \label{eq:R_m}
\end{align}
and $\mar_\ee$ is the achievable rate over the eavesdropper channel given by
\begin{align}
\mar_\ee=\log\abs{\mI+\rho_\ee\mH_\ee \mQ \mH_\ee^\her} \label{eq:R_e}
\end{align} 
with $\mQ_{N_t \times N_t}$ being the diagonal power allocation~matrix.~Co- nsidering the \ac{tas} protocol $\mas$, $\mQ$ reads
\begin{align}
[\mQ]_{ww} = \left\{
        \begin{array}{ll}
            1 & w\in \setW_\mas \\
            0 & w \notin \setW_\mas
        \end{array}
    \right.
\end{align}
in which the average transmit power on each selected antenna is set to be one. Consequently, $\mar_\mm$ and $\mar_\ee$ reduce to
\begin{subequations}
\begin{align}
\mar_\mm &=\hspace*{.1mm}\log\abs{\mI+\rho_\mm\tmH_\mm^\her \tmH_{\mm}} \label{eq:R_mS} \\
\mar_\ee &=\log\hspace*{.1mm}\abs{\mI+\rho_\ee\tmH_\ee^\her \tmH_{\ee}} \label{eq:R_eS}
\end{align}
\end{subequations}
where $\tmH_\mm$ and $\tmH_\ee$ are $N_\rr \times L_\rmt$ matrices denoting the effective main and eavesdropper channels respectively. The effective channels are constructed from $\mH_\mm$ and $\mH_\ee$ by collecting the columns which correspond to the selected antennas. Substituting in \eqref{eq:sys-R_s}, the maximum~achievable~secrecy rate reads
\begin{align}
\mar_\rms\prnt{\mas}=\left[ \log\frac{\abs{\mI+{\rho_m}\tmH_\mm^\her \tmH_\mm}}{\abs{\mI+{\rho_\ee}\tmH_\ee^\her \tmH_\ee}}\right]^+  \label{eq:R_sS}
\end{align}
where the argument $\mas$ is written to indicate the dependency of $\mar_\rms\prnt{\mas}$ on the \ac{tas} protocol.

Using $\mar_\rms$ in \eqref{eq:sys-R_s}, different secrecy measures for the system is defined based on the eavesdropper's status. When the \ac{csi} of the eavesdropper channel is available at the transmit side, \eqref{eq:sys-R_s} determines the instantaneous achievable secrecy rate, and thus, its expected value determines the ergodic secrecy rate. For a passive eavesdropper, the probability of $\mar_\rms$ being less than $\mar_\rmO$ evaluates the outage probability which represents probability of having information leakage when the transmitter has set the secrecy rate to $\mar_\rmO$ \cite{barros2006secrecy}.
%

\section{Large-System Secrecy Performance of \ac{tas}}
\label{sec:result}
In this section, we investigate the secrecy performance of the \ac{mimome} wiretap channel under the \ac{tas} protocol~$\mas$~considering the following two cases:
\begin{enumerate}[label= \Alph*]
\item \label{itmA} The eavesdropper is equipped with significantly fewer receive antennas compared to the number of~selected~antennas, i.e., $N_\ee \ll L_\rmt$.
\item \label{itmB} The number of eavesdropper antennas grows large faster than the number of selected antennas, i.e., $N_\ee \gg L_\rmt$.
\end{enumerate}
Case \ref{itmA} can be seen as a scenario in cellular networks with the eavesdropper being a regular user terminal. Moreover, Case \ref{itmB} describes a scenario in which the eavesdropper is a sophisticated terminal, such as portable stations.

Proposition \ref{thm:1} approximates the distribution of the maximum achievable secrecy rate in the large-system limit, i.e., $N_\rmt\uparrow\infty$, for both Cases \ref{itmA} and \ref{itmB}.
\begin{proposition}
\label{thm:1}
Consider the \ac{tas} protocol $\mas$, and let 
\begin{subequations}
\begin{align}
\eta_t&=N_\mathrm{r}\left[ L_\mathrm{t}  + N_\mathrm{t} f_{N_\mathrm{r} +1}(u) \right] \label{eq:eta_t} \\
\sigma_t^2&=\left(L_\mathrm{t}u-\eta_t \right)^2 \left(\frac{1}{L_\mathrm{t}}-\frac{1}{N_\mathrm{t}} \right) - \frac{\eta_t^2}{L_\mathrm{t}}+ \Xi_t \label{eq:sigma_t}
\end{align}
\end{subequations}
for some non-negative real $u$ which satisfies
\begin{align}
\int_u^\infty f_{N_\mathrm{r}}(x) \mathrm{d} x= \frac{L_\mathrm{t}}{N_\mathrm{t}},
\end{align}
$\Xi_t$ which is given by
\begin{align}
\Xi_t\coloneqq N_\mathrm{r} \left(N_\mathrm{r}+1\right) \left[ L_\mathrm{t}+N_\mathrm{t} f_{N_\mathrm{r}+1}(u)+N_\mathrm{t} f_{N_\mathrm{r}+2}(u) \right],
\end{align}
and $f_{N_\mathrm{r}}(\cdot)$ which represents the chi-square probability density function with $2N_\mathrm{r}$ degrees of freedom and mean $N_\mathrm{r}$, i.e.,
 \begin{equation}
 \label{eq:Je}
    f_{N_\mathrm{r}}(x)= \frac{1}{(N_\mathrm{r}-1)!}
    \begin{cases}
     x^{N_\mathrm{r}-1} e^{-x} , & \text{if}\ x \geq 0 \\
      0, & \text{if}\ x < 0.
    \end{cases}
  \end{equation}
Define $L_\mm$ $\coloneqq\min\set{L_\mathrm{t},N_\mathrm{r}}$, $M_\mm\coloneqq\max\set{L_\mathrm{t},N_\mathrm{r}}$, $L_\ee\coloneqq\min\set{L_\mathrm{t},N_\mathrm{e}}$ and $M_\ee\coloneqq\max\set{L_\mathrm{t},N_\mathrm{e}}$. As $N_\rmt$ grows large, the distribution of $\mar_\rms(\mas)$ for both Cases \ref{itmA} and \ref{itmB} can be approximated by the distribution of $\mar_\mathrm{asy}(\mas)\coloneqq\left[ \mar^\star\right]^+$ in which $\mar^\star$ is a Gaussian random variable with mean $\eta$ and variance $\sigma^2$ given by \eqref{eq:eta_final} and \eqref{eq:sigma_final} on the top of the~next~page.
\end{proposition}
\begin{figure*}[!t]

\begin{align}
\eta&\coloneqq L_\mm \log \left( 1+ \frac{\rho_\mm \eta_t}{L_\mm}\right) - \dfrac{L_\mm\left(L_\mm-1 \right)\rho_\mm^2\eta_t^2 }{2M_\mm\left( L_\mm + \rho_\mm \eta_t \right)^2} \log e -L_\ee \log \left( 1+ \rho_\ee M_\ee \right) \label{eq:eta_final} 
\end{align}
\begin{align}
\sigma^2 &\coloneqq \left( \left[ \frac{ L_\mm \rho_\mm}{L_\mm+\rho_\mm \eta_t}-\dfrac{ L_\mm^2 \left( L_\mm-1 \right)\rho_\mm^2 \eta_t }{ M_\mm \left( L_\mm + \rho_\mm \eta_t \right)^3} \right]^2 \sigma_t^2 +\mone_{\set{N_\ee > L_\rmt}}\frac{L_\ee}{M_\ee} +\mone_{\set{N_\ee < L_\rmt}} \frac{L_\ee M_\ee \rho^2_\ee}{\left(1+\rho_\ee M_\ee\right)^2  } \right) \log^2 e \label{eq:sigma_final}
\end{align}
\centering\rule{17cm}{0.1pt}
%
\vspace*{2pt}
\end{figure*}
\begin{prf}
To evaluate the large-system distribution of $\mar_\rms(\mas)$, one needs to determine the asymptotic distribution of $\mar_\mm$ and $\mar_\ee$ given by \eqref{eq:R_m} and \eqref{eq:R_e} respectively. Using the results from \cite{asaad2017tas} and \cite{hochwald2004multiple}, the distribution of $\mar_\mm$ and $\mar_\ee$ in the large system limits for the both cases \ref{itmA} and \ref{itmB} can be approximated by Gaussian distributions. Noting the fact that the main and eavesdropper channels are independent, the random variable
\begin{align}
\mar^\star \coloneqq \mar_\mm-\mar_\ee, \label{eq:R_star}
\end{align}
in the large limit, can then be approximated with a Gaussian random variable whose variance and mean is determined in terms of the variances and means of $\mar_\mm$ and $\mar_\ee$. Finally by substituting in \eqref{eq:sys-R_s}, the proof is concluded. The detailed derivations are given in the appendix.
\end{prf}

Considering \eqref{eq:sigma_final}, one can observe that the variations of the secrecy rate vanish as the dimensions of the system increase. In fact, as $N_\rmt$ grows large $\eta_t$ grows proportionally, and therefore, the first term in \eqref{eq:sigma_final} tends to zero. Moreover, as $L_\ee / M_\ee$ in our setup is considered to be significantly small, the two other terms can be neglected further. Consequently, in the~large~limit $\sigma$ converges to zero; the observation which could be intuitively predicted, due to the fact that the both main and eavesdropper channels harden in the large limit. The mean value $\eta$, however, does not necessarily increase in the large-system limit, since it is given as the difference of two terms which can both asymptotically grow large. The latter observation indicates that increasing the number of selected antennas for this setup does not necessarily improve the secrecy rate. We discuss this argument later in Section \ref{sec:sec_enh}. At this point, we employ Proposition \ref{thm:1} to evaluate the ergodic secrecy rate and secrecy outage probability.
\subsection{Ergodic Secrecy Rate}
\begin{figure}[t]
\hspace*{-.82cm}  
\resizebox{1.06\linewidth}{!}{
\pstool[width=.35\linewidth]{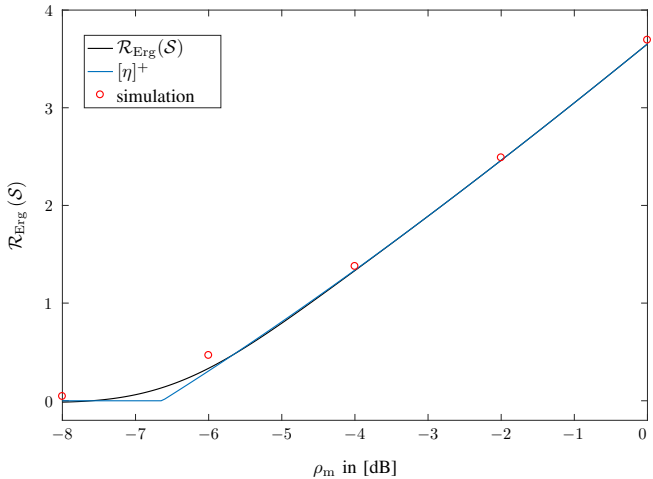}{
\psfrag{Rerg}[c][t][0.25]{$\mar_\rmE\prnt{\mas}$}
\psfrag{Lt}[c][c][0.25]{$\rho_\mm$ in [dB]}
\psfrag{RR-NN-Nr=2}[l][l][0.25]{$\mar_\rmE(\mas)$}
\psfrag{EE-NN-Nr=2}[l][l][0.25]{$[\eta]^+$}
\psfrag{SS-NN-Nr=2}[l][l][0.25]{simulation}

\psfrag{-8}[c][c][0.22]{$-8$}
\psfrag{-7}[c][c][0.22]{$-7$}
\psfrag{-6}[c][c][0.22]{$-6$}
\psfrag{-5}[c][c][0.22]{$-5$}
\psfrag{-4}[c][c][0.22]{$-4$}
\psfrag{-3}[c][c][0.22]{$-3$}
\psfrag{-2}[c][c][0.22]{$-2$}
\psfrag{-1}[c][c][0.22]{$-1$}
\psfrag{0}[c][c][0.22]{$0$}

\psfrag{0}[c][c][0.22]{$0\hspace*{2mm}$}
\psfrag{1}[c][c][0.22]{$1\hspace*{2mm}$}
\psfrag{2}[c][c][0.22]{$2\hspace*{2mm}$}
\psfrag{3}[c][c][0.22]{$3\hspace*{2mm}$}
\psfrag{4}[c][c][0.22]{$4\hspace*{2mm}$}
\psfrag{5}[c][c][0.22]{$5\hspace*{2mm}$}
}}
\caption{The ergodic secrecy rate in terms of the main channel's \ac{snr}. The curves have been plotted for $N_\rr=N_\ee=2$, $L_\mt=8$, $N_\rmt=16$ and $\rho_\ee=-5$ dB. $\mar_\rmE(\mas)$ in \eqref{eq:asy} and the numerical simulations are sketched by a black line and red circles respectively. The blue line indicate $[\eta]^+$ for $\eta$ given in \eqref{eq:eta_final}. As it shows, the approximation tracks the simulations with high accuracy even for finite dimensions.\vspace*{-2mm}}
\label{fig:1}
\end{figure}

For scenarios in which the \ac{csi} of the eavesdropper channel is available at the transmit side, the instantaneous secrecy rate is achievable in each transmission interval. Therefore, the secrecy performance is characterized by the ergodic secrecy rate which is given by taking the expectation of $\mar_\rms\prnt{\mas}$. Using Proposition \ref{thm:1}, the ergodic secrecy rate for our setup in~the~large limit is approximated as
\begin{subequations}
\begin{align}
\mar_\rmE \prnt{\mas} &\approx \E \set{\mar_\asy\prnt{\mas}}\\
&=\E \set{\left[\mar^\star\right]^+}\\
&= \sigma \hspace*{.5mm} \phi\prnt{\xi}+ \eta \hspace*{.5mm}\rmQ\prnt{-\xi}. \label{eq:asy}
\end{align}
\end{subequations}
where $\xi\coloneqq\dfrac{\eta}{\sigma}$. Using the inequality $\rmQ(x) < \dfrac{\phi\prnt{x}}{x}$ for $x>0$, we conclude
\begin{align}
\mar_\rmE\prnt{\mas} > \eta \label{eq:lower}
\end{align}
for $\xi>0$; the bound is tight when $\xi$ is large enough. Fig.~\ref{fig:1} illustrates the accuracy of the approximations, as well as the tightness of the lower-bound in \eqref{eq:lower}. The figure has been plotted for $N_\rr=N_\ee=2$ and $L_\rmt=8$ considering $16$ transmit antennas and $\rho_\ee=-5$ dB. As the figure shows, except for the interval of $\rho_\mm$ in which $\eta$ is close to zero, the lower-bound in \eqref{eq:lower} perfectly matches $\mar_\rmE(\mas)$. This observation is due to fact that the variance in the large limit tends to zero rapidly, and thus, the factor $\xi$ grows significantly large even for finite values of $\eta$; therefore, $\mQ(-\xi)\approx1-\xi^{-1} \phi(\xi)$, and the ergodic secrecy rate is approximated with $\eta$ accurately. Despite the initial assumptions on the system dimensions taken in Sections \ref{sec:sys} and \ref{sec:result}, one observes that the approximation accurately tracks the simulations even in not so large dimensions.

\subsection{Secrecy Outage Probability}
When the eavesdropper is passively overhearing the channel, the \ac{csi} of the eavesdropper channel is not known at the transmitter, and thus, the secrecy rate in \eqref{eq:sys-R_s} is not achievable. The performance in this case is described by the secrecy outage probability which for a given rate $\mar_\rmO\geq 0$~is~defined~as~\cite{barros2006secrecy}
\begin{align}
\Pout_\rmO\prnt{\mar_\rmO} = \Pr\set{\mar_s \prnt{\mas} < \mar_\rmO }. \label{eq:secrecy_outage}
\end{align}
The interpretation of this secrecy measure is as follows: As the transmitter does not have the \ac{csi} of the eavesdropper~channel, it sets the secrecy rate to be $\mar_\rmO$ in all transmission intervals. This setting implicitly imposes a primary assumption on the quality of the eavesdropper channel. In this~case,~the outage probability in \eqref{eq:secrecy_outage} determines the probability of eavesdropper channel having better quality than the primary assumption, or equivalently, the percentage of intervals in which the eavesdropper can decode transmitted codewords at~least~partially.
 
Using the large-system approximation in Proposition \ref{thm:1}, the outage probability in the large limit reads
\begin{subequations}
\begin{align}
\Pout_\rmO\prnt{\mar_\rmO} &\approx \Pr\set{\mar_\asy \prnt{\mas} \leq \mar_\rmO }\\
&= 1- \rmQ\prnt{\frac{\mar_\rmO - \eta}{\sigma}}. \label{eq:out}
\end{align}
\end{subequations}

\begin{figure*}
$\qquad$\vspace*{-6pt}
%
\end{figure*}
\section{Secrecy Enhancement via \ac{tas}}
\label{sec:sec_enh}
Considering either the ergodic secrecy rate or the secrecy outage probability, the secrecy performance of the system in the large limit is mainly specified by $\eta$. In contrast to $\sigma$ which tends to zero in the asymptotic regime, the factor $\eta$, for a given number of receive and eavesdropper antennas, can either grow, vanish or tend to some constant in the large-system limit depending on the number of selected antennas. 

\begin{figure}[t]
\hspace*{-.82cm}  
\resizebox{1.06\linewidth}{!}{
\pstool[width=.35\linewidth]{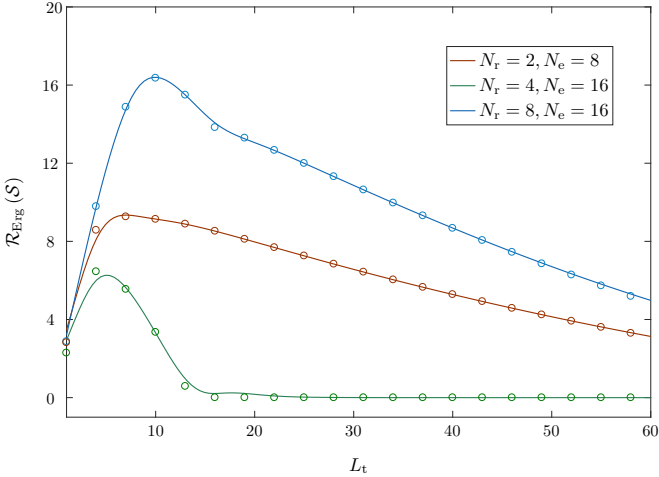}{
\psfrag{Rerg}[c][t][0.25]{$\mar_\rmE\prnt{\mas}$}
\psfrag{Lt}[c][c][0.25]{$L_\rmt$}
\psfrag{Nr=08-Ne=16-Nt}[l][l][0.25]{$N_\rr=8, N_\ee=16$}
\psfrag{Nr=02-Ne=08-Nt}[l][l][0.25]{$N_\rr=2, N_\ee=8$}
\psfrag{Nr=04-Ne=16-Nt}[l][l][0.25]{$N_\rr=4, N_\ee=16$}

\psfrag{10}[c][c][0.22]{$10$}
\psfrag{20}[c][c][0.22]{$20$}
\psfrag{30}[c][c][0.22]{$30$}
\psfrag{40}[c][c][0.22]{$40$}
\psfrag{50}[c][c][0.22]{$50$}
\psfrag{60}[c][c][0.22]{$60$}

\psfrag{0}[c][c][0.22]{$0\hspace*{3mm}$}
\psfrag{4}[c][c][0.22]{$4\hspace*{3mm}$}
\psfrag{8}[c][c][0.22]{$8\hspace*{3mm}$}
\psfrag{12}[c][c][0.22]{$12\hspace*{2mm}$}
\psfrag{16}[c][c][0.22]{$16\hspace*{2mm}$}
\psfrag{18}[c][c][0.22]{$20\hspace*{2mm}$}
}}
\caption{The ergodic secrecy rate as a function of $L_\rmt$ for three different cases considering $N_\rmt=128$, $\rho_\mm=0$ dB and $\rho_\ee=-10$ dB. The approximation in \eqref{eq:asy} and numerical simulations are denoted by solid lines and circles respectively. As it shows, the values of $L_\rmt$ in which $\mar_\rmE(\mas)$ meets the maximum is significantly smaller than~$N_\rmt$.}
\label{fig:3}
\end{figure}

To illustrate the point further, we have plotted in Fig.~\ref{fig:3} the ergodic secrecy rate of the system as a function of $L_\rmt$, for some given numbers of receive and eavesdropper antennas considering both the large-system approximation given by \eqref{eq:asy}, and numerical simulations. The curves have been sketched for $\rho_\mm=0$ dB and $\rho_\ee=-10$ dB. As the figure shows, for the considered setups, the ergodic secrecy rate meets its maximum at some values of $L_\rmt$ which are significantly smaller than $N_\rmt$. The observation which expresses that the \ac{tas} in these scenarios, not only benefits in terms of \ac{rf}-cost reduction, but also enhances the secrecy performance of the system. The intuition behind this behavior comes from the fact that the growth in the number of selected antennas improves both the main and eavesdropper channels. In this case as Fig.~\ref{fig:3} indicates, the improvement from the legitimate receiver's point of view dominates the overall secrecy performance of the system up to a certain number of selected antennas. By increasing $L_\rmt$ further, the improvement at the eavesdropper side starts to dominate which results in the performance deficit at larger values of $L_\rmt$.

\subsection{Optimal Number of Selected Antennas}
Considering the illustrated behavior of the system, a smart choice of $L_\rmt$ can significantly improve the overall system throughput at no cost. In this case, the results given in Section \ref{sec:result} can be employed, in order to find the optimal number of selected transmit antennas. More precisely, using \eqref{eq:eta_final} and \eqref{eq:sigma_final}, one can determine the ergodic secrecy rate and secrecy outage probability in \eqref{eq:asy} and \eqref{eq:out} as functions of $L_\rmt$ whose maximizers are found either analytic or via a linear search. We investigate this problem further through the following examples by considering the ergodic secrecy rate as the measure. Same discussions can also be considered for the secrecy outage probability

\begin{example}[Single-antenna receivers]
\label{ex:1}
Consider the case in which the legitimate receiver, as well as the eavesdropper, is equipped with a single receive antenna, i.e., $N_\rr=N_\ee=1$. This can be seen as a scenario in which both the legitimate receiver and eavesdropper are handheld devices. Considering the ergodic secrecy rate as the performance measure,~we~are interested in determining the optimal number of~selected~antennas which maximizes $\mar_\rmE\prnt{\mas}$.

Let us denote the optimal number of selected antennas by $L^\star_\rmt$. In order to employ the results in Section \ref{sec:result}, we assume at this point that $L_\rmt^\star \gg 1$; we later show that the assumption holds for the large limit of $N_\rmt$. By substituting in \eqref{eq:eta_final} and \eqref{eq:sigma_final}, $\eta$ and $\sigma^2$ are determined as
\begin{subequations}
\begin{align}
\eta&=\log \prnt{\frac{1+\rho_\mm L_\rmt\prnt{1+ \loge{N_\rmt L_\rmt^{-1}}}}{1+\rho_\ee L_\rmt}}\\
\sigma^2&= \left[ \frac{\rho_\mm^2 \hspace*{1mm} L_\rmt \left( 2- L_\rmt N_\rmt^{-1} \right)}{\prnt{1+\rho_\ee L_\rmt}^2} + \frac{L_\rmt \rho_\ee^2}{(1+\rho_\ee L_\rmt)^2} \right] \log^2 e.
\end{align}
\end{subequations}
By taking the limit $N_\rmt\uparrow\infty$ and considering $L^\star_\rmt\gg 1$, it is concluded that $\xi$ is large, and therefore, $\mar_\rmE\prnt{\mas}\approx \eta$. To find $L^\star_\rmt$, we define the function
\begin{align}
f(x) \coloneqq \log \prnt{\frac{1+\rho_\mm x+\rho_\mm x \hspace*{.5mm} \loge{N_\rmt x^{-1}}}{1+\rho_\ee x}}
\end{align}
over the real axis. It is then straightforward to show that, for $x\in[1,N_\rmt]$, $f''(x) \leq 0$, and thus, one can conclude that $L^\star_\rmt$ is an integer close to the maximizer of $f(\cdot)$. Consequently, $L^\star_\rmt$ can be approximated as $L^\star_\rmt\approx \lfloor x^\star\rceil$ where $x^\star$ satisfies
\begin{align}
\rho_\ee x^\star + \loge {x^\star} + \frac{\rho_\ee}{\rho_\mm}= \loge{N_\rmt}. \label{eq:L_opt}
\end{align}
As $x^\star$ grows large proportional to $N_\rmt$, $L_\rmt^\star$ grows accordingly, and therefore, the initial assumption of $L_\rmt^\star \gg 1$ in the large-system limit holds. Moreover, by reducing $\rho_\ee$ to zero in the fixed point equation \eqref{eq:L_opt}, $L_\rmt^\star=N_\rmt$ which agrees with the fact that in the absence of the eavesdropper, the ergodic rate is a monotonically increasing function of $L_\rmt$.

\begin{figure}[t]
\hspace*{-.82cm}  
\resizebox{1.06\linewidth}{!}{
\pstool[width=.35\linewidth]{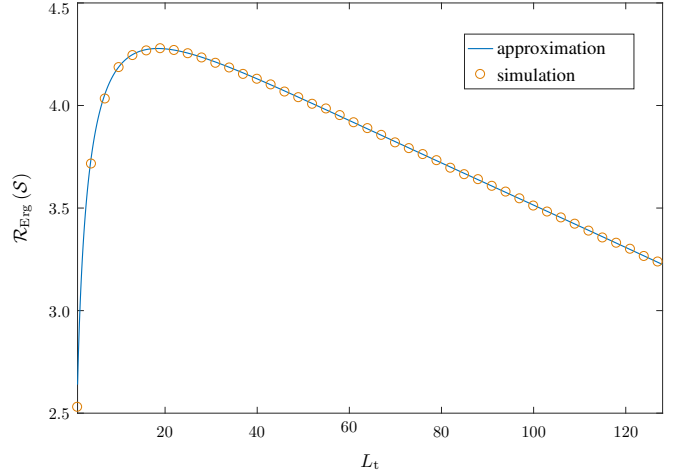}{
\psfrag{Rerg}[c][t][0.25]{$\mar_\rmE\prnt{\mas}$}
\psfrag{Lt}[c][c][0.25]{$L_\rmt$}
\psfrag{Nr=1-Ne=1-Nt=S}[l][l][0.25]{approximation}
\psfrag{Nr=1-Ne=1-Nt=R}[l][l][0.25]{simulation}

\psfrag{20}[c][b][0.22]{$20$}
\psfrag{40}[c][b][0.22]{$40$}
\psfrag{60}[c][c][0.22]{$60$}
\psfrag{80}[c][b][0.22]{$80$}
\psfrag{100}[c][b][0.22]{$100$}
\psfrag{120}[c][b][0.22]{$120$}

\psfrag{2.5}[c][c][0.22]{$2.5\hspace*{2mm}$}
\psfrag{3}[c][c][0.22]{$3.0\hspace*{4mm}$}
\psfrag{3.5}[c][c][0.22]{$3.5\hspace*{2mm}$}
\psfrag{4}[c][c][0.22]{$4.0\hspace*{4mm}$}
\psfrag{4.5}[c][b][0.22]{$4.5\hspace*{2mm}$}
}}
\caption{$\mar_\rmE(\mas)$ in Example~\ref{ex:1} in terms of the number of selected antennas. The solid line and the circles show the approximation given by \eqref{eq:asy} and numerical simulations respectively considering $N_\rmt=128$, $\rho_\mm=0$ dB and $\rho_\ee=-10$ dB. As it is observed, $L_\rmt^\star=18$ is suggested by both the approximation and simulation results.}
\label{fig:4}
\end{figure}

Fig.~\ref{fig:4} shows the ergodic secrecy rate as a function of $L_\rmt$ for $\rho_\ee=-10$ dB and $128$ transmit antennas. By solving the fixed point equation in \eqref{eq:L_opt}, $x^\star=18.4$ is obtained which results in $L_\rmt^\star=18$. The result which is confirmed by numerical simulations as well.
\end{example}

\begin{figure*}
$\qquad$\vspace*{-6pt}
%
\end{figure*}

\begin{figure}[t]
\hspace*{-.82cm}  
\resizebox{1.06\linewidth}{!}{
\pstool[width=.35\linewidth]{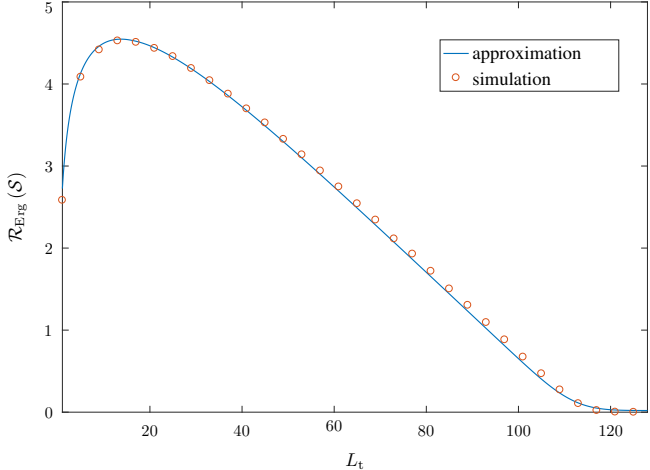}{
\psfrag{Rerg}[c][t][0.25]{$\mar_\rmE\prnt{\mas}$}
\psfrag{Lt}[c][c][0.25]{$L_\rmt$}
\psfrag{Nr=1-Ne=16-Nt=S}[l][l][0.25]{approximation}
\psfrag{Nr=1-Ne=16-Nt=R}[l][l][0.25]{simulation}

\psfrag{20}[c][b][0.22]{$20$}
\psfrag{40}[c][b][0.22]{$40$}
\psfrag{60}[c][c][0.22]{$60$}
\psfrag{80}[c][b][0.22]{$80$}
\psfrag{100}[c][b][0.22]{$100$}
\psfrag{120}[c][b][0.22]{$120$}

\psfrag{0}[c][c][0.22]{$0\hspace*{2mm}$}
\psfrag{1}[c][c][0.22]{$1\hspace*{2mm}$}
\psfrag{2}[c][c][0.22]{$2\hspace*{2mm}$}
\psfrag{3}[c][c][0.22]{$3\hspace*{2mm}$}
\psfrag{4}[c][c][0.22]{$4\hspace*{2mm}$}
\psfrag{5}[c][b][0.22]{$5\hspace*{2mm}$}
}}
\caption{The ergodic secrecy rate in Example~\ref{ex:2} vs. $L_\rmt$ for the case with $N_\ee=16$, $N_\rmt=128$, $\rho_\mm=0$ dB and $\rho_\ee=-25$ dB. The large-system approximation and numerical simulations are denoted by the solid line and circles respectively, and both suggest $L_\rmt^\star=14$.}
\label{fig:5}
\end{figure}

\begin{example}[Multi-antenna eavesdropper]
\label{ex:2}
As another example, we consider a scenario with a single antenna legitimate receiver whose channel is being overheard by a sophisticated muti-antenna terminal, i.e., $N_\rr=1$ and $N_\ee$ growing large. Similar to Example \ref{ex:1}, let us take the ergodic secrecy rate as the performance measure. For this case, $\eta$ and $\sigma^2$ in \eqref{eq:eta_final} and \eqref{eq:sigma_final} reduce to
\begin{subequations}
\begin{align}
\eta&=\log \prnt{\frac{1+\rho_\mm L_\rmt\prnt{1+ \loge{N_\rmt L_\rmt^{-1}}}}{(1+\rho_\ee N_\ee)^{L_\rmt}}} \label{eq:ex-21} \\
\sigma^2&= \frac{\rho_\mm^2 \log^2 e \hspace*{1mm} L_\rmt \left( 2- L_\rmt N_\rmt^{-1} \right)}{\prnt{1+\rho_\ee L_\rmt}^2} + \frac{\log^2 e}{L_\rmt}. \label{eq:ex-22}
\end{align}
\end{subequations}
In contrast to Example \ref{ex:1}, the ratio $\xi$ in this case~does~not~necessarily take large values, and consequently, $\mar_\rmE\prnt{\mas}$ is not approximated by $\eta$. Therefore, we define $c(\cdot)$ over $\setR$ as
\begin{align}
c(x) \coloneqq s(x)\phi\prnt{h(x)} + f(x) \rmQ\prnt{-h(x)}
\end{align}
where $f(x)$ and $[s(x)]^2$ are obtained by replacing $L_\rmt$ with $x$ in \eqref{eq:ex-21} and \eqref{eq:ex-22} respectively, and $h(x)\coloneqq [s(x)]^{-1}f(x)$. With similar lines~of~inference the optimal number of selected antennas is approximated as $L^\star_\rmt\approx \lfloor x^\star\rceil$ with $x^\star$ satisfying
\begin{align}
\hspace*{-2mm}\phi\prnt{h(x^\star)} \set{2f(x^\star)h'(x^\star)\hspace*{-.7mm}-\hspace*{-.7mm}s'(x^\star)}\hspace*{-1mm} = \hspace*{-1mm} f'(x^\star) \rmQ\prnt{-h(x^\star)}. \label{eq:L2}
\end{align}
In Fig.~\ref{fig:5}, $\mar_\rmE(\mas)$ is sketched in terms of the number of selected antennas for $N_\ee=16$, $\rho_\ee=-25$ dB and $128$ transmit antennas. From the fixed point equation in \eqref{eq:L2}, the extreme point of the curve occurs at $x^\star=13.7$ which recovers $L_\rmt^\star=14$ given by the simulation results.
\end{example}

\section{Conclusion}
\label{conclusion}
In this paper, we have characterized the effect of multiple \ac{tas} on the secrecy performance of a massive \ac{mimome} channel by considering the ergodic secrecy rate and secrecy outage probability. For this setup, the instantaneous secrecy rate has been approximated in terms of a Gaussian random variable when the number of transmit antennas grows large. This approximation enabled us to show that, for several~scenarios, the secrecy performance of the system is maximized at some optimal number of selected antennas. Our numerical simulations showed that the optimal number of transmit antennas is accurately approximated for finite dimensions. 

\appendices
\section*{Appendix: Derivation of Proposition \ref{thm:1}}
We start by determining the large-system distribution of $\mar_\mm$ given in \eqref{eq:R_m}. In \cite[Lemma~2]{asaad2017tas}, it has been shown~that the distribution of the input-output mutual information of a Gaussian \ac{mimo} channel, under some constraints, is approximated sufficiently close to its exact value in terms of the random variables $\tr{\mJ}$ and $\tr{\mJ^2}$ where $\mJ\coloneqq\mH^\her \mH$. Under the \ac{tas} protocol $\mas$, $\tr{\mJ}$ represents the sum of $L_\rmt$ first order statistics which at the large limit of $N_t$ converges in distribution to a Gaussian random variable whose mean and variance are given by \eqref{eq:eta_t} and \eqref{eq:sigma_t}, respectively \cite{asaad2017tas}. Using some properties of random matrices, the large-system distribution of $\mar_\mm$ is then approximated as in \cite[Proposition~1]{asaad2017tas} with a Gaussian distribution whose mean and variance are given in terms of $\eta_t$ and $\sigma_t^2$.

The next step is to evaluate the distribution of $\mar_\ee$. Noting that the main and eavesdropper channels are independent, it is concluded that the \ac{tas} protocol $\mas$ performs as a random selection protocol from the eavesdropper's point of view. Therefore, $\tr{\mJ}$ and $\tr{\mJ^2}$ in both the cases \ref{itmA} and \ref{itmB} can be determined explicitly as functions of independent Gaussian random variables. By substituting in \cite[Lemma~1]{asaad2017tas} and taking the limit of $N_\rmt$ growing large, the large-system distribution of $\mar_\ee$ can be approximated with a Gaussian distribution whose mean and variance are respectively given by
\begin{subequations}
\begin{align}
\hspace*{-2mm} \eta_\ee &= L_\ee \log\prnt{1+\rho_\ee M_\ee}\\
\hspace*{-2mm}\sigma_\ee^2 &= \left( \mone_{\set{N_\ee > L_\rmt}}\frac{L_\ee}{M_\ee}\hspace*{-.7mm} +\hspace*{-.7mm}\mone_{\set{N_\ee < L_\rmt}} \frac{L_\ee M_\ee \rho^2_\ee}{\left(1+\rho_\ee M_\ee\right)^2  } \right) \log^2 e
\end{align}
\end{subequations}
which recovers the asymptotic results for \ac{iid} Gaussian fading channels given in \cite{hochwald2004multiple}. Using the independency of the main and eavesdropper channels, $\mar^\star$ in \eqref{eq:R_star} can be approximated in the large system limit with a Gaussian~random~variable whose mean and variance are as in \eqref{eq:eta_final}~and~\eqref{eq:sigma_final}.


\bibliography{ref}
\bibliographystyle{IEEEtran}

\begin{acronym}
\acro{mimo}[MIMO]{Multiple-Input Multiple-Output}
\acro{mimome}[MIMOME]{Multiple-Input Multiple-Output Multiple-Eavesdropper}
\acro{csi}[CSI]{Channel State Information}
\acro{awgn}[AWGN]{Additive White Gaussian Noise}
\acro{iid}[i.i.d.]{independent and identically distributed}
\acro{ut}[UT]{User Terminal}
\acro{bs}[BS]{Base Station}
\acro{tas}[TAS]{Transmit Antenna Selection}
\acro{lse}[LSE]{Least Squared Error}
\acro{rhs}[r.h.s.]{right hand side}
\acro{lhs}[l.h.s.]{left hand side}
\acro{wrt}[w.r.t.]{with respect to}
\acro{rs}[RS]{Replica Symmetry}
\acro{rsb}[RSB]{Replica Symmetry Breaking}
\acro{papr}[PAPR]{Peak-to-Average Power Ratio}
\acro{rzf}[RZF]{Regularized Zero Forcing}
\acro{snr}[SNR]{Signal-to-Noise Ratio}
\acro{rf}[RF]{Radio Frequency}
\acro{mf}[MF]{Match Filtering}
\end{acronym}

\usepackage [active,tightpage]{preview}
\let \document \origdocument 
\pagestyle {empty}
\begin {document}
\centering \null \vfill 
\begin {preview}
 \psfrag {Pr}[c][c][0.25]{$12$} 
\includegraphics [width=.35\linewidth ] {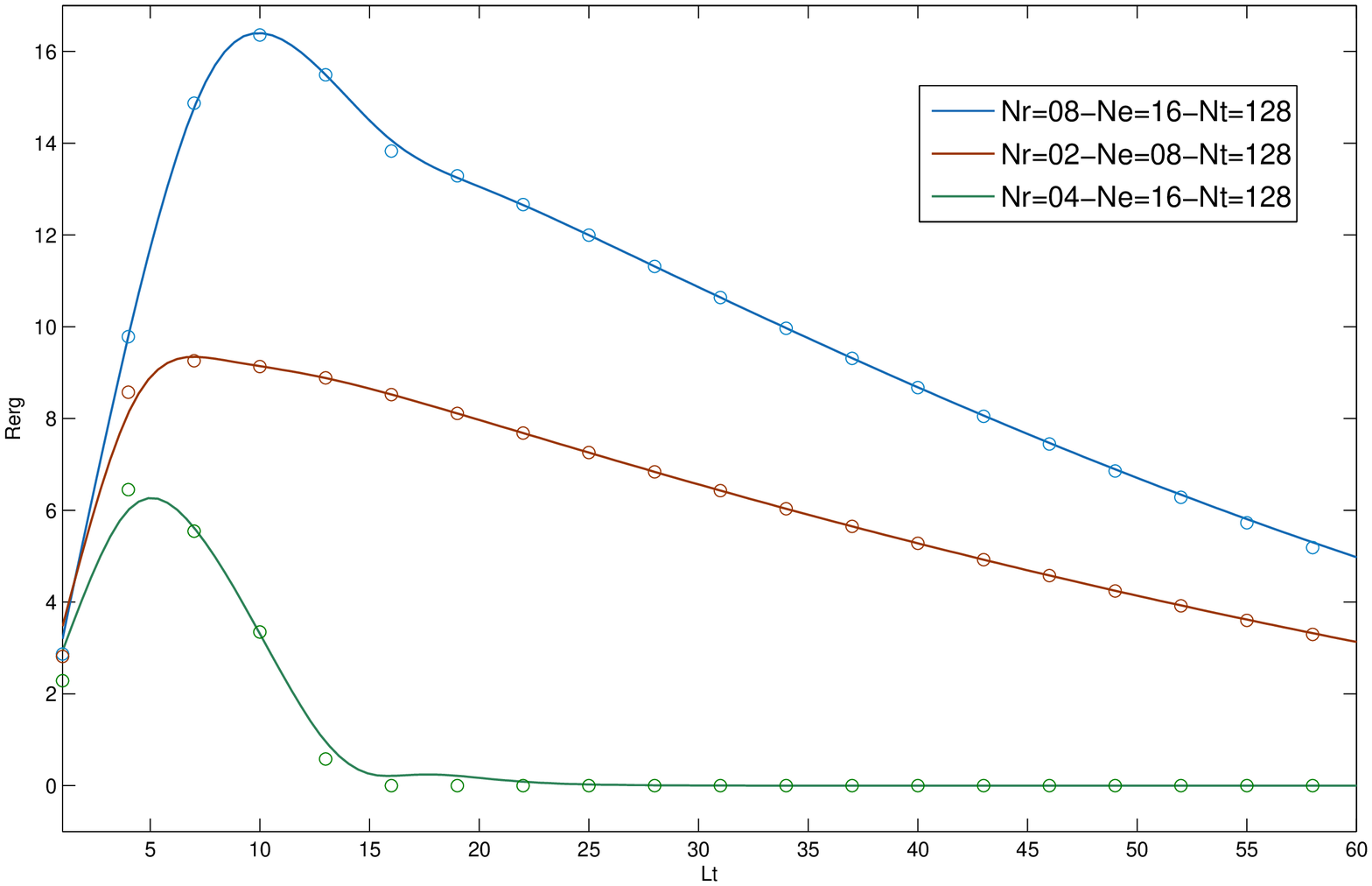}
\end {preview}
\vfill \end {document}